\documentclass[aps,prd,showpacs,notitlepage,nofootinbib,superscriptaddress,floatfix,showkeys,twocolumn]{revtex4-2}

\usepackage[parfill]{parskip}    
\usepackage{graphicx}
\usepackage{subfigure}
\usepackage{amssymb}
\usepackage{bm}
\usepackage{epstopdf} 
\usepackage{amsmath}
\usepackage{mathrsfs}
\usepackage{varioref}
\usepackage{mathtools}
\usepackage{hyperref} 

\hypersetup{
	bookmarks=false,         
	pdfstartview={FitH},    
	colorlinks=true,       
	linkcolor=cyan,          
	citecolor=green,        
	filecolor=blue,      
	urlcolor=blue           
}

\DeclareFontFamily{OT1}{pzc}{}
\DeclareFontShape{OT1}{pzc}{m}{it}%
{<-> s * [0.900] pzcmi7t}{}
\DeclareMathAlphabet{\mathscr}{OT1}{pzc}%
{m}{it}
\labelformat{section}{Section #1} 
\labelformat{subsection}{Section #1} 
\labelformat{equation}{Eq.~(#1)} 
\labelformat{figure}{Fig.~#1} 
\labelformat{subfigure}{Fig.~\thefigure#1} 
\labelformat{table}{Tab.~#1} 
\newcommand{\be}{\begin{equation}}
	\newcommand{\ee}{\end{equation}}
\newcommand{\bea}{\begin{eqnarray}}
	\newcommand{\eea}{\end{eqnarray}}

\def\Mpl{M_{_{\rm Pl}}}

\def\f{\frac}
\def\l{\left}
\def\r{\right}

\begin{document}
\title{Conserved cosmological perturbations in USR inflation and bouncing scenarios}
 \author{Rathul Nath Raveendran}
 \email{rathulnath.r@gmail.com}
 \affiliation{ School of Physical Sciences, Indian Association for the Cultivation of Science, Kolkata~700032, India}

\begin{abstract}
Inflationary and bouncing scenarios are two frameworks that provide the  mechanism to overcome the
horizon problem as well as generate the primordial perturbations. In this work, we investigate the conservation of perturbations in single-field models of both inflationary and bouncing scenarios, where the quantity, $z = a \, \rm d \phi/{\rm d}\log a$, with $a$ representing the scale factor and $\phi$ denoting the scalar field, decreases with time. We observe that this behaviour occurs during the ultra-slow-roll phase in the context of inflation and the contracting phase in the context of bounce. We show that the conjugate momentum associated with the comoving curvature perturbation during both the ultra-slow-roll phase and the contracting phase of bouncing scenarios is conserved in the super-Hubble limit. We illustrate that, within the framework of inflation, this conservation of momentum allows for the evolution of perturbations across the ultra-slow-roll phase, enabling the calculation of the power spectrum for modes that exit the Hubble radius before the ultra-slow-roll phase begins. Similarly, in the context of a bounce, we can determine the power spectrum after the bounce using this method. We support our approach with both numerical and analytical arguments.
\end{abstract}
\maketitle
\flushbottom
\section{Introduction}
The primordial curvature power spectrum, which leads to the anisotropies in the Cosmic Microwave Background (CMB) and later to the distribution of galaxies, is constrained by precise cosmological observations on large scales~\cite{Ade:2015lrj,Planck:2018jri}. The spectrum of perturbations on large scales is a key
test of various models of the early universe.
Among various scenarios considered, the inflationary paradigm stands out as the most popular model for explaining the origin of perturbations in the early universe. According to the theory of inflation, the quantum fluctuations of the scalar fields during
the early universe are responsible for the primordial perturbations. In the standard scenario, the inflationary phase is driven by one or more scalar fields
which slowly roll along a nearly flat potential. It is important to emphasize that the simplest models of slow-roll inflation, where all the slow-roll parameters are small, predict a nearly scale-invariant spectrum that aligns with observations of the CMB~\cite{Mukhanov:1988jd, Mukhanov:1990me, Martin:2013tda}.  

However, it has been known that, the nearly scale-invariant power spectrum can also arise in another class of models known as ultra-slow-roll inflation. In these models, the first slow-roll parameter is exceptionally small, but the second one exceeds one. However, it is crucial to highlight that ultra-slow-roll inflation encounters numerous challenges, as the system proves to be unstable and requires fine-tuned initial conditions~\cite{Inoue:2001zt, Martin:2012pe, Namjoo:2012aa,Pattison:2018bct}. 

In recent times, there has been a growing interest in exploring ultra-slow-roll inflation. This interest arises from the realization that an ultra-slow-roll phase towards the end of inflation can result in the production of a substantial number of primordial black holes. It has been known that during inflation, a considerable boost in the primordial scalar power on small scales, compared to the COBE normalized amplitude over large scales, can lead to the generation of primordial black holes. In the case of single field models of inflation, it has
been found that, a point of inflection in the potential leads to an epoch of ultra-slow-roll, which can give rise to the required boost in the scalar amplitude over
small scales~\cite{Garcia-Bellido:2017mdw, Germani:2017bcs,Ballesteros:2017fsr,Ragavendra:2020sop,Ragavendra:2023ret,Jackson:2023obv,Choudhury:2013woa}.

Despite the remarkable success of inflation, alternative scenarios have been explored to explain the origin of primordial perturbations. The most extensively investigated alternative to the inflationary paradigm is the classical non-singular bouncing scenario. In a non-singular bouncing scenario, the universe experiences a period of contraction until the scale factor reaches a minimum value, after which it transitions to a phase of expansion. The Hubble parameter is negative during the contraction phase and positive during the expansion phase. These two phases are connected by a bouncing phase in which the rate of change of the Hubble parameter is positive~\cite{Khoury:2001wf,Buchbinder:2007ad,Lehners:2007ac, Brandenberger:2012aj, Brandenberger:2016vhg, Raveendran:2017vfx}. 

In single field models of inflationary and bouncing scenarios, the evolution of perturbations can be obtained by understanding the evolution of the scale factor and scalar field. More precisely, the quantity $z = a\,{\rm d}\phi/{\rm d}\log a$, where $a$ represents the scale factor and $\phi$ denotes the scalar field, serves as the background quantity governing the evolution of perturbations. The quantity $z$ is sometimes described as the pump field for scalar perturbations. Interestingly, in slow-roll inflation, it is known that the quantity $z$ increases, and the comoving curvature perturbation ${\cal R}$ is often conserved in the super Hubble limit. However, in ultra-slow-roll inflation and the contracting phases of bouncing scenarios, $z$ decreases, resulting in the growing nature of ${\cal R}$ in these scenarios.

 In this work, we investigate the conserved quantity during the ultra-slow-roll and contracting phases of bouncing scenarios on super-Hubble scales, exploring its applications. We demonstrate that the conserved quantity is the conjugate momentum associated with the comoving curvature perturbation. Furthermore, after establishing the conservation of momentum in these phases both analytically and numerically, we discuss its applications in realistic models that incorporate these phases.

This article is organized as follows: In \ref{sec:action}, we provide a concise overview of perturbations and the corresponding equations derived from the action. Proceeding to \ref{sec:conservation}, we examine the conservation perturbations in the super-Hubble limit. \ref{sec:pl-scalef} is dedicated to investigating the conservation of perturbations in the context of power-law inflation and the contracting phase of bouncing scenarios. In \ref{sec:USR} and \ref{sec:bounce}, we extend our analysis to more realistic models of ultra-slow-roll inflation and bouncing scenarios. Finally, a succinct conclusion is presented in \ref{sec:conclusion}.

Conventions and notations: we will utilize natural units, setting $\hbar$ and $c$ to 1, while defining the Planck mass as $\Mpl=(8 \pi G)^{-1/2}$. Consistent with standard practice, differentiation with respect to the cosmic time $t$ will be indicated by a dot, while differentiation with respect to the conformal time $\eta$ will be indicated by a prime. 


\section{Evolution of perturbations} \label{sec:action}
In the case of early universe models driven by a single, canonical scalar field, the
scalar perturbations are governed by a single quantity, comoving curvature perturbation, that is often denoted as ${\cal R}$. 
At the linear order in perturbation theory, the comoving curvature perturbation is governed by the following action~\cite{Mukhanov:1990me}:
\begin{equation}
    A= \frac{1}{2}\int {\textrm d} \eta \, {\textrm d} x^3\,  z^2 \left( {\cal R}'^2  - \left(\Delta {\cal R}\right)^2 \right)\,,
\end{equation}
where $z=a\, \dot{\phi}/H$, with $\phi$ being the scalar field which drives the background evolution. The conjugate momentum associated with ${\cal R}$ can be obtained as
\begin{equation}
    \Pi = z^2\,{\cal R}'\,.
\end{equation}
Then the Hamiltonian can be constructed as
\begin{equation}
    H = \frac{1}{2} \int {\textrm d} x^3 \left( z^{-2} \, \Pi^2  + z^2 \, \left(\Delta {\cal R}\right)^2\right)\,, \label{eq:H}
\end{equation}
and the corresponding Hamilton’s equations of motion are given by 
\begin{equation}
    {\cal R}' = \frac{\Pi}{z^2}\, , \quad \Pi' = z^2 \, \left(\Delta {\cal R}\right)^2 \,.
\end{equation}
It is useful to note that, these equations lead to two decoupled second-order equations as 
\begin{subequations}
\begin{eqnarray}
    {\cal R}'' + 2 \frac{z'}{z} {\cal R}' - \Delta^2 {\cal R} &=&0\,, \\
    \Pi'' - 2 \frac{z'}{z} \, \Pi' - \Delta^2 \, \Pi &=& 0\, . 
\end{eqnarray}\label{eq:H-eqns-real-space}
\end{subequations}

In Fourier space, we find that the Hamilton's equations~\ref{eq:H-eqns-real-space} can be written as
\begin{equation}
    {\cal R}_{\bm k}' = \frac{\Pi_{\bm k}}{z^2}\, , \quad \Pi_{\bm k}' = -z^2 \, k^2 {\cal R}_{\bm k} \,,\label{eq:H-eqns-k}
\end{equation}
where ${\cal R}_{\bm k}$ and $\Pi_{\bm k}$ are the Fourier modes associated with the comoving curvature perturbation and the corresponding momentum. 
These equations lead to two decoupled second-order equations
\begin{subequations}
\begin{eqnarray}
 \label{eq:eom-R}   {\cal R}'' + 2 \frac{z'}{z} {\cal R}' + k^2 {\cal R} &=&0\, , \\
    \Pi'' - 2 \frac{z'}{z} \, \Pi' + k^2 \, \Pi &=& 0\, , 
\end{eqnarray}\label{eq:eom-R-Pi}
\end{subequations}
where we have dropped the subscript ${\bm k}$ for convenience.

It is interesting to note that the transformations
\begin{equation}
    \Pi\rightarrow \Tilde{\Pi} = k\, {\cal R}\, , \, {\cal R} \rightarrow \Tilde{{\cal R}} = - k^{-1} \Pi\, , \, z^2 \rightarrow \Tilde{z}^2 = z^{-2}\,,
\end{equation}
leaves the Hamiltonian (\ref{eq:H}) and equations of motion (\ref{eq:H-eqns-k}) unchanged~\cite{Brustein:1998kq,Boyle:2004gv}. In terms of Mukhanov-Sasaki variables the above equations can be rewritten as
\begin{subequations}
\begin{eqnarray}
    v'' + \left(k^2-\frac{z''}{z} \right) v &=&0\, , \\
    p'' + \left(k^2-\frac{\theta''}{\theta} \right) p &=&0\,,
\end{eqnarray}\label{eq: eom-v-p}
\end{subequations}
where $v= z {\cal R}$, $p = \Pi/z$ and $\theta = 1/z$. It is useful to write $z''/z$ and $\theta''/\theta$ in terms of slow-roll parameters as 
\begin{subequations}
\begin{align}
\hskip -40pt \frac{1}{\left(a H\right)^2}\frac{z''}{z}&=  2- \epsilon_1 + \frac{3}{2}\epsilon_2 -\frac{1}{2}\epsilon_1 \epsilon_2 + \frac{1}{4} \epsilon_2^2 +\frac{1}{2} \epsilon_2 \epsilon_3\,,\\
\frac{1}{\left(a H\right)^2}\frac{\theta''}{\theta}&= \epsilon_1 + \frac{1}{2}\epsilon_2 +\frac{1}{2}\epsilon_1 \epsilon_2 + \frac{1}{4} \epsilon_2^2 -\frac{1}{2} \epsilon_2 \epsilon_3\,, \label{eq:zppz-tppt}
\end{align}
\end{subequations}
where the slow-roll parameters are defined as
\begin{equation}\label{eq:def-epsilons}
    \epsilon_1= - \frac{{\rm d\, log}H}{{\rm d} N}\,, \quad
    \epsilon_{n+1}= \frac{{\rm d\, log}\epsilon_n}{{\rm d} N}\,.
\end{equation}
The quantity of observational significance is the power spectrum of comoving curvature perturbation, denoted as $\mathcal{P}_{\mathcal{R}}$. Which is conventionally defined as
\begin{equation}
\mathcal{P}_{\mathcal{R}} = \frac{k^3}{2\pi^2} \left| {\cal R}\right|^2\,,
\end{equation}
where ${\cal R}$ is obtained using \ref{eq:eom-R-Pi} by imposing Bunch-Davies initial conditions \cite{Mukhanov:1990me}.
\section{Conserved quantities}\label{sec:conservation}
It has been established that in the context of inflation, when the scales of cosmological interest are significantly outside the horizon, certain perturbations can be conserved under appropriate conditions. This property often eliminates the need for more detailed information about the background. One such perturbation is the comoving curvature perturbation, denoted as ${\cal R}$. On the super-Hubble limit, where $k^2 {\cal R} \rightarrow 0$, this conservation can be seen from~\ref{eq:eom-R-Pi}. Under this limit,~\ref{eq:eom-R-Pi} leads to
\begin{equation}
{\cal R}' (\eta) \propto \frac{1}{z^2(\eta)}\,.
\end{equation}
The expression above clearly indicates that $\cal R' $ decreases with an increase in $z^2$. Consequently, it implies that $\cal R$ remains conserved when $z^2$ increases, as observed in the context of slow-roll inflation.

However, in the case of USR inflation, one can observe that the quantity
\begin{equation}
\frac{z'}{z} = a\, H \left(1+\frac{\epsilon_2}{2}\right)\,,
\end{equation}
can be negative since $\epsilon < -2$. This implies that $z^2$ decreases during the USR phase. Consequently, the conservation of the superhorizon curvature perturbation, is no longer valid as $z^2$ decreases in this case. Nevertheless, when examining the evolution of $\Pi$ on the super-Hubble limit, we obtain
\begin{equation}
{\Pi}' (\eta)\propto z^2(\eta)\,.\label{eq:Pip-approx}
\end{equation}
This implies that $\Pi'$ diminishes as $z^2$ decreases, suggesting that $\Pi$ is conserved when $z^2$ decreases, as observed in USR inflation. It is important to note that, in the evolution of momentum, the super-Hubble limit is defined by $k^2\ll \theta''/\theta$, whereas in the case of curvature perturbation, the limit is characterized by $k^2\ll z''/z$, as indicated by the  \ref{eq: eom-v-p}.

Additionally, it is interesting to note that in the case of a contracting phase, the Hubble parameter is negative, and $\epsilon_2$ is either positive or negligible. Consequently, $z^2$ decreases. This suggests that the quantity $\Pi$ is conserved during the contracting phase as well.
\section{Power law scale factor}\label{sec:pl-scalef}
Let us now investigate the conservation of the mentioned quantities in the last section under different background evolutions of our interest. First, we consider a power-law form for the scale factor
\begin{equation}\label{eq:a-pl}
    a(\eta) = \bar{a} \left(\frac{\eta}{\bar{\eta}}\right)^{\frac{1}{\epsilon-1}}\,,
\end{equation}
where $\bar{a}, \bar{\eta}$ and $\epsilon$ are constants. In this case, it is known that the equations governing the perturbations (\ref{eq:eom-R-Pi}) can be solved analytically when Bunch-Davies initial conditions are assumed. These solutions can be written in terms of Hankel functions  as
\begin{subequations}
			\begin{align}			
			{\cal R}(\eta)&=\frac{1}{a \sqrt{2 \epsilon}\,\Mpl}\,\sqrt{\frac{-\eta\,  \pi}{4}}\,e^{i (\nu +1/2)\pi/2} H_{\nu}^{(1)}(-k \eta)\, ,\\
			\Pi(\eta)&=-a \sqrt{2 \epsilon}\, \Mpl \sqrt{\frac{-\eta\,  \pi}{4}}\,ke^{i (\nu +1/2)\pi/2}  H_{\nu-1}^{(1)}(-k \eta)\,,
			\end{align}\label{eq:R-Pi-Hankel}
		\end{subequations}
where $\nu \equiv \frac{1}{2} + \frac{1}{1-\epsilon}$. 
We find that, on super-Hubble limit (when $-k\,\eta \ll 1$), the limiting form of the Hankel function can be written as
\begin{subequations}
\begin{align}
H_{\nu}^{(1)}(-k \eta) &\simeq -\frac{i}{\pi}\Gamma(\nu)\left(\frac{-k\eta}{2}\right)^{-\nu}~\,\, {\textrm for}\,\nu>0 \,,\\ 
H_{\nu}^{(1)}(-k \eta) &\simeq -i\,\frac{e^{-i\nu \pi}}{\pi}\Gamma(-\nu)\left(\frac{-k\eta}{2}\right)^{\nu} \,\, {\textrm for}\, \nu<0 \, .\label{eq:lim_Hank}
\end{align}
\end{subequations}
\subsection{Expanding phase}
During inflationary expansion, where $\epsilon<1$ on super-Hubble limit, the equations \ref{eq:R-Pi-Hankel} and \ref{eq:lim_Hank} lead to
\begin{subequations}
\begin{align}
k^{3/2}\vert {\cal R} (\eta) \vert & \approx \frac{k}{2\,a \, \sqrt{\pi}\, \sqrt{\epsilon}\,\Mpl}\,\Gamma(\nu) \left(\frac{-k\, \eta}{2}\right)^{1/2-\nu} \propto \eta^0\,,\\
\frac{\Pi (\eta)}{k^{3/2}} &\approx\frac{a\, \sqrt{\epsilon}\,\Mpl}{k \sqrt{\pi}} \, \Gamma(\nu-1) \left(\frac{-k\,\eta}{2}\right)^{3/2-\nu} \propto \eta^{2(1-\nu)}\,.
\end{align}
\end{subequations}
The above equations clearly illustrate that the comoving curvature perturbation is conserved on the super-Hubble limit in power-law inflation, while its associated conjugate momentum increases. 
\subsection{Contracting phase}
Now, let's examine the evolutions of ${\cal R}$ and $\Pi$ during the power-law contracting phase. Consider the scenario where ${1 < \epsilon < 3}$. This encompasses cases where the contracting phase is dominated by matter ({\it i.e.,} 
 when $\epsilon = 3/2$) \cite{Wands:1998yp}. In this context, one can obtain
\begin{subequations}
\begin{align}
k^{3/2}\vert {\cal R} (\eta) \vert & \approx \frac{k}{2\,a \, \sqrt{\pi}\, \sqrt{\epsilon} \,\Mpl}\,\Gamma(\nu) \left(\frac{-k\, \eta}{2}\right)^{1/2+\nu} \propto \eta^{2 \nu}\,,\\
\frac{\Pi (\eta)}{k^{3/2}} &\approx\frac{a\, \sqrt{\epsilon}\, \Mpl}{k \sqrt{\pi}} \, \Gamma(1-\nu) \left(\frac{-k\,\eta}{2}\right)^{\nu-1/2} \propto \eta^0\,.
\end{align}
\end{subequations}
It is evident from the above expressions that the conjugate momentum $\Pi$ is conserved, as expected from the expression in \ref{eq:Pip-approx}. Moreover, in the scenario of a matter-dominated contracting phase, where $\epsilon = 3/2$, the expressions above yield a scale-invariant spectrum for ${\cal R}$~\cite{Wands:1998yp}.

Another interesting scenario occurs when ${\epsilon > 3}$, commonly referred to as the ekpyrotic contracting phase~\cite{Khoury:2001wf}. In this case
\begin{subequations}
\begin{align}
k^{3/2}\vert {\cal R} (\eta) \vert & \approx \frac{k}{2\,a \, \sqrt{\pi}\, \sqrt{\epsilon}\, \Mpl}\,\Gamma(\nu) \left(\frac{-k\, \eta}{2}\right)^{1/2-\nu}  \propto \eta^0\,,\\
\frac{\Pi (\eta)}{k^{3/2}} &\approx \frac{a\, \sqrt{\epsilon}\, \Mpl}{k \sqrt{\pi}} \, \Gamma(1-\nu) \left(\frac{-k\,\eta}{2}\right)^{\nu-1/2}  \propto \eta^0\,.
\end{align}
\end{subequations}

 indicate that not only is the conjugate momentum conserved, but the comoving curvature perturbation is also conserved.

 The above discussions highlight two significant outcomes. First, it reaffirms the well-known result from existing literature that the quantity ${\cal R}$ is conserved on super-Hubble scales in an expanding universe. Secondly, and more importantly in the context of our analysis, it showcases that the quantity $\Pi$ remains conserved on super-Hubble scales during the contracting phase, as realized in the~\ref{sec:conservation}.
\section{Ultra-Slow-Roll inflation}\label{sec:USR}
Let us now explore scenarios where both decreases and increases in $z^2$ occur during the evolutions. First, we consider inflation, characterized by three periods of evolution: slow-roll, followed by ultra-slow-roll, and then again slow-roll. In the first slow-roll phase, $z^2$ increases, then decreases during the ultra-slow-roll phase, and increases once more during the subsequent slow-roll phase. Another context is bouncing scenarios, where $z^2$ decreases during contraction and increases during expansion.

First, we examine inflation, specifically a model with a brief ultra-slow-roll phase. To achieve this, we consider a model described by the potential~\cite{Dalianis:2018frf,Ragavendra:2020sop}
\begin{align}
V(\phi) = V_0\,&\biggl\{\mathrm{tanh}\l(\f{\phi}{\sqrt{6}\,\Mpl}\r)
 \nonumber\\
&+ A\,\sin\l[\f{1}{f_\phi}\,
\mathrm{tanh}\l(\f{\phi}{\sqrt{6}\,\Mpl}\r)\r]\biggr\}^2.
\label{eq:usr}
\end{align}
The potential features a point of inflection, and we opt to work with the following parameter values: $V_0 = 2 \times 10^{-10}\,\Mpl^4$, $A = 0.130383$, and $f_\phi = 0.129576$. With these parameter values, the point of inflection in the potential is situated at $\phi_0 = 1.05\,\Mpl$~\cite{Ragavendra:2020sop}. If we set the initial field value to $\phi_i=6.1\,\Mpl$, with $\epsilon_{1i}=10^{-4}$, we achieve approximately $66$~e-folds of inflation in the model. Additionally, we assume that the pivot scale exits the Hubble radius about $56.2$~e-folds prior to the termination of inflation.
\begin{figure}[!t]
\centering
\includegraphics[width=0.9\linewidth]{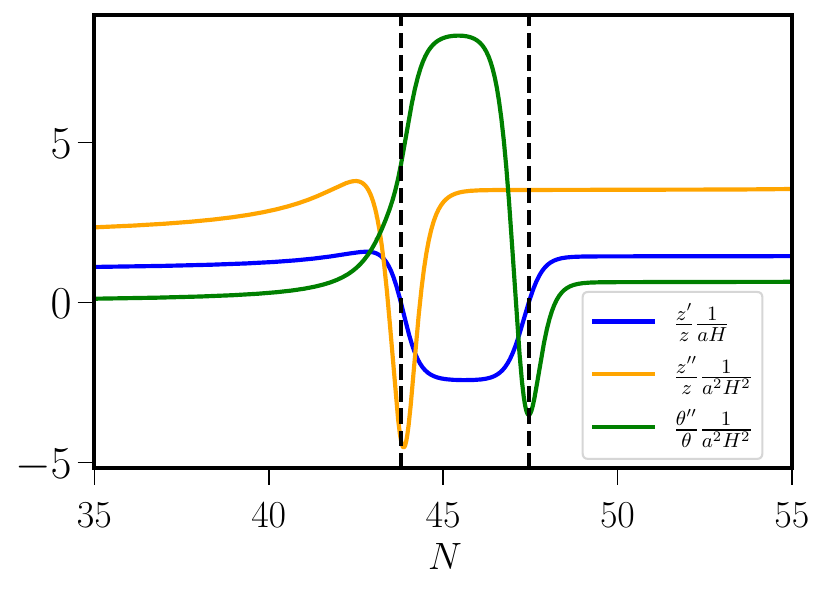}
\caption{Evolutions of $z'/z$, $z''/z$, and $\theta''/\theta$. Vertical lines are positioned at the values of $N$ corresponding to $43.789$ and $47.456$, marking the interval where $z^2$ decreases. }\label{fig:zppz-usr}
\end{figure}

\begin{figure*}[!t]
\centering
\includegraphics[width=0.475\linewidth]{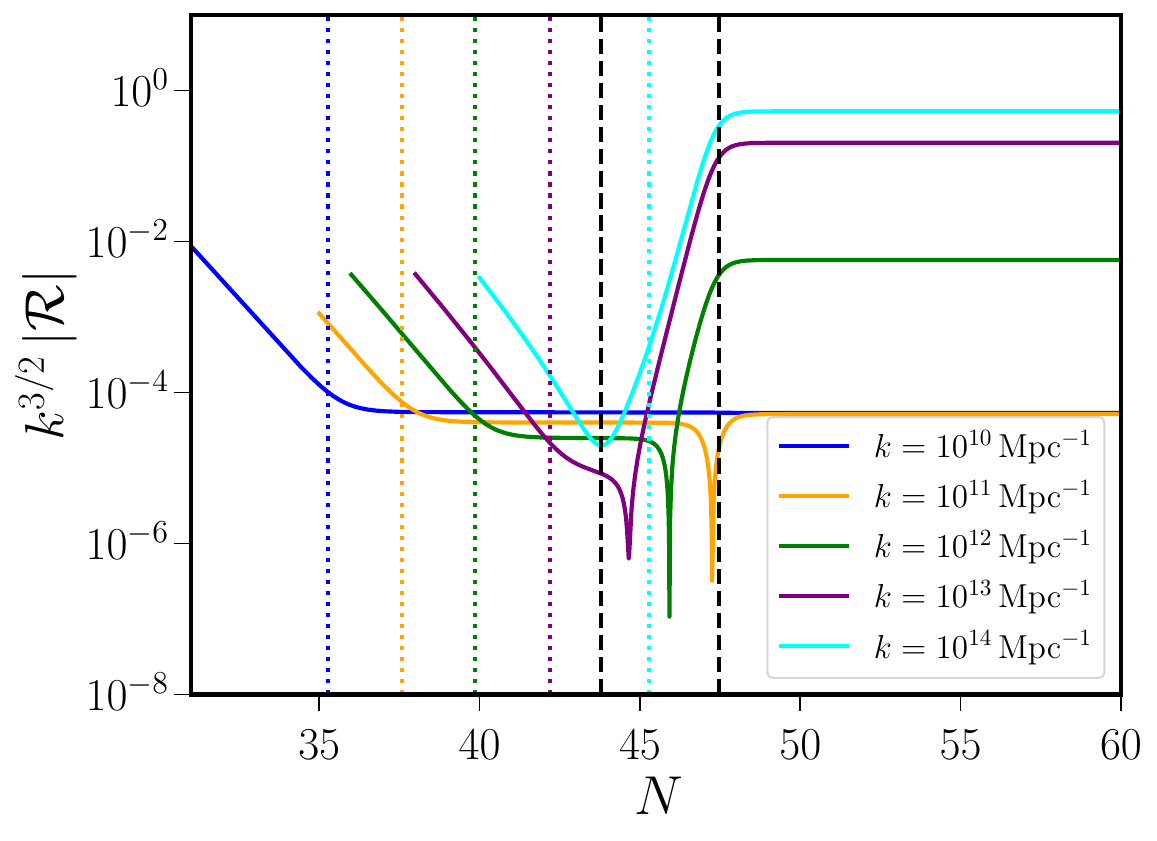}
\hskip 5pt
\includegraphics[width=0.475\linewidth]{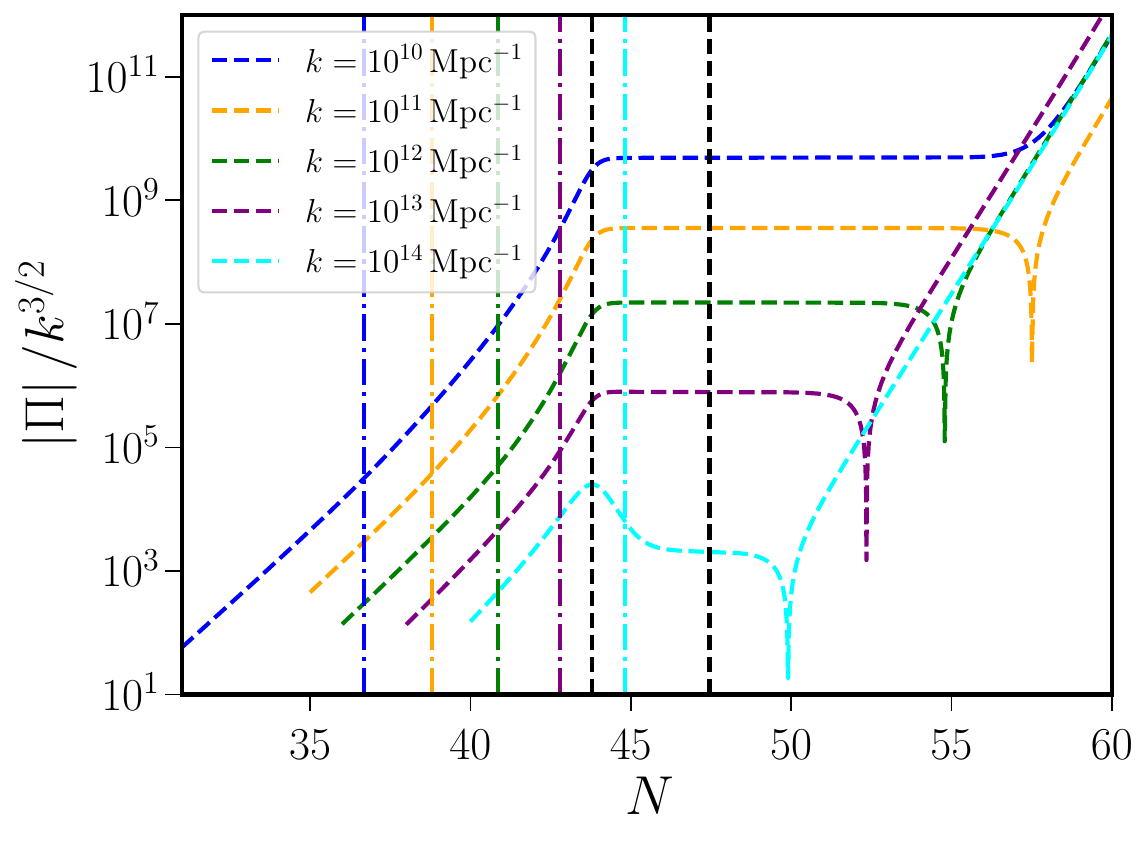}
\caption{Evolutions of $k^{3/2}\vert{\cal R}\vert$ and $\vert \Pi\vert/k^{3/2}$ as functions of $N$ for $k$ values of $10^{10}$, $10^{11}$, $10^{12}$, $10^{13}$ and $10^{14} \, {\rm Mpc^{-1}}$. Dotted vertical lines and dashed-dotted lines mark the instances where $z''/z = k^2$ and $\theta''/\theta = k^2$, respectively, for $k= 10^{10}$ (blue), $k= 10^{11}$ (orange), $k= 10^{12}$ (green), $k= 10^{13}$ (purple), and $k= 10^{14}\, {\rm Mpc^{-1}}$ (cyan). Dashed vertical lines are positioned at $N$ values corresponding to $43.789$ and $47.456$, marking the interval during which $z^2$ decreases. These figures demonstrate the conservation of ${\cal R}$ during the slow-roll phase, as well as the conservation of $\Pi$ during the phase where $z^2$ decreases.}\label{fig:R-P-evolution}
\end{figure*}
\subsection{Analytical expressions}
In the case of ultra-slow-roll inflation, it has been known that an accurate analytical expression can be derived from \ref{eq:eom-R} by performing an expansion in $k^2$ [see Refs. \cite{Leach:2001zf, Jackson:2023obv} in this context]. Our next task is to investigate whether the conservation of ${\cal R}$ and $\Pi$ can be utilized to rederive this analytical expression for the power spectrum at the end of inflation. To achieve this, we focus on the modes that exit the Hubble radius well before the ultra-slow-roll phase. For these modes, we know that ${\cal R}$ is conserved on the super-Hubble limit during the slow-roll phase. Utilizing this information and solving the \ref{eq:H-eqns-k}, we obtain
\begin{eqnarray}
    \Pi(\eta_\ast) &=& \Pi(\eta_{\rm i})-k^2 {\cal R}(\eta_{\rm i}) \int_{\eta_{\rm i}}^{\eta_\ast} z^2 {\textrm d} \eta\,.\label{eq:Pi-approx}
\end{eqnarray}
Here, $\eta_{\rm i}$ is selected such that the mode is in the super-Hubble limit, and $\eta_\ast$ represents a time slightly after the mode enters the phase where $z^2$ decreases. Subsequently, given the conservation of $\Pi$ during this phase, we derive
\begin{eqnarray}
    {\cal R}(\eta_{\rm f}) &=& {\cal R}(\eta_{\rm i}) + \Pi(\eta_\ast) \int_{\eta_\ast}^{\eta_{\rm f}} \frac{1}{z^2}  {\textrm d} \eta\,,\label{eq:R-approx}
\end{eqnarray}
where $\eta_{\rm f}$ represents the time slightly after the phase where $z^2$ decreases. It is important to emphasize that, in deriving the expressions in ~\ref{eq:R-approx} and \ref{eq:Pi-approx}, we make the assumption that the conservation of ${\cal R}$ extends until $\eta_\ast$, despite this being the time within the phase of decreasing $z^2$. Similarly, we also assume the conservation of $\Pi$ until $\eta_{\rm f}$, even though $\eta_{\rm f}$ is outside this phase. For numerical calculations, it is useful to express the expressions~\ref{eq:Pi-approx} and \ref{eq:R-approx} in terms of the number of e-folds, as
    \begin{eqnarray}
  \Pi(N_\ast) &=& \Pi(N_{\rm i})-k^2 {\cal R}(N_{\rm i}) \int_{N_{\rm i}}^{N_\ast} \frac{z^2}{a H} {\textrm d} N\,,\\
    {\cal R}(N_{\rm f}) &=& {\cal R}(N_{\rm i}) + \Pi(N_\ast) \int_{N_\ast}^{N_{\rm f}} \frac{1}{z^2 a H}  {\textrm d} N\,. \label{eq:R-Pi-approx-N}
\end{eqnarray}
As previously mentioned, an analytical expression for ${\cal R}$ can be derived from the integral solution of \ref{eq:eom-R} for small but finite wavenumbers, up to order $O(k^2)$ \cite{Leach:2001zf, Jackson:2023obv}. Interestingly, one can see that the expressions \ref{eq:Pi-approx} and \ref{eq:R-approx} derived by using the conservation of ${\cal R}$ and $\Pi$ are consistent with these existing results.

\ref{fig:zppz-usr} depicts the evolutions of $z'/z$, $z''/z$, and $\theta''/\theta$ as function of number of e-folds $N$. Note that vertical dashed lines mark the interval during which $z^2$ decreases.~\ref{fig:R-P-evolution} illustrates the evolution of $\vert{\cal R}\vert$ and $\vert \Pi\vert$ as a function of $N$. The figures clearly demonstrate the conservation of $\Pi$ during the phase where $z^2$ decreases on super-Hubble limit. Also, note the amplification in the evolution of $\vert{\cal R}\vert$ for the modes that exit the Hubble radius closer to the ultra-slow-roll phase. We find that, this amplification occurs when the second term becomes dominant in~\ref{eq:R-approx} due to the decreasing nature of $z^2$. Furthermore, we observe that the evolution of the conjugate momentum is smooth during the ultra-slow-roll phase without any dip, unlike the evolutions of ${\cal R}$ for certain wavenumbers.

\begin{figure}[!t]
\centering
\includegraphics[width=0.9\linewidth]{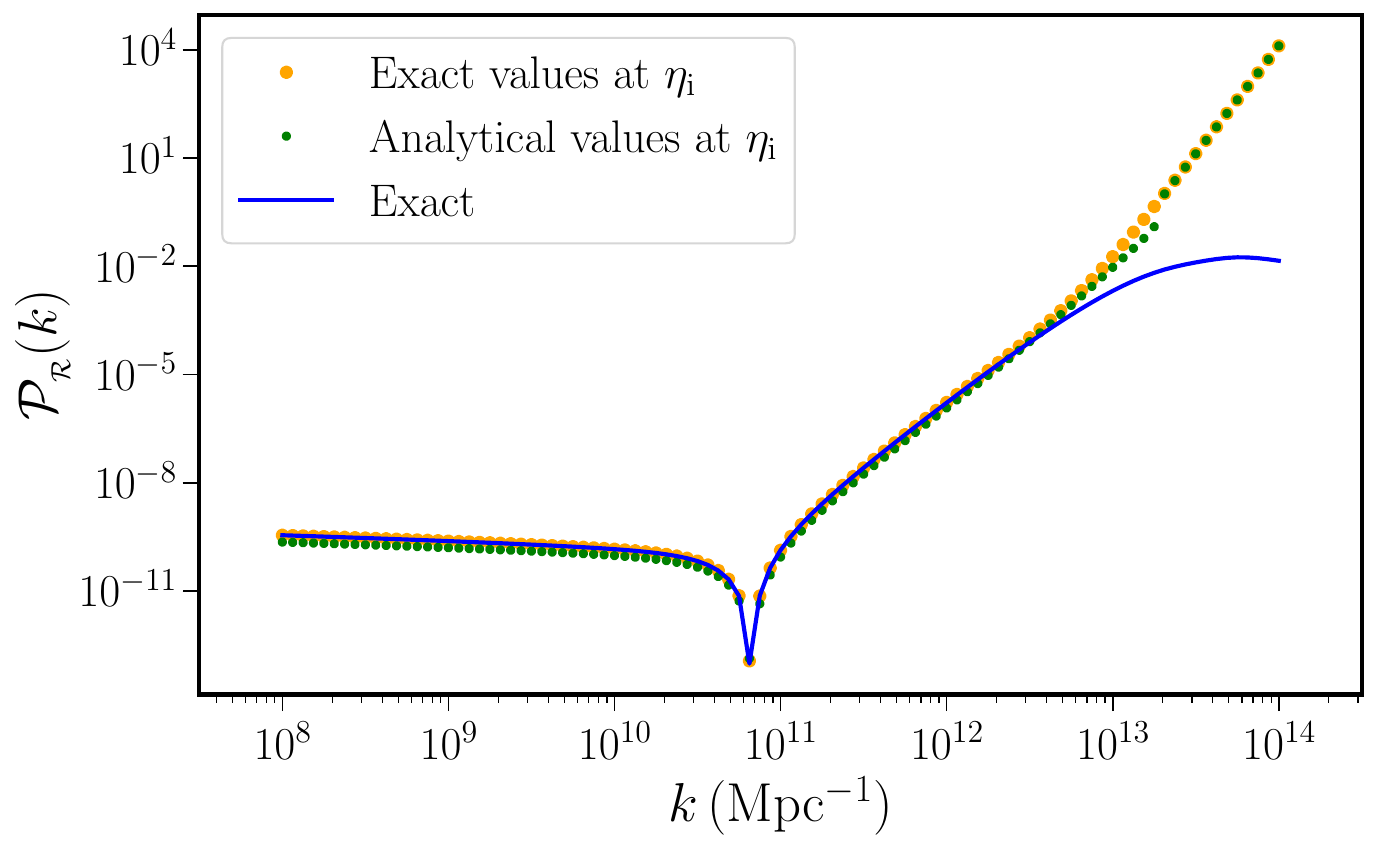}
\caption{Power spectrum of comoving curvature perturbations. The blue line represents the numerically calculated power spectrum by solving the \ref{eq:H-eqns-k}. The orange data points illustrate the spectrum calculated using~\ref{eq:R-Pi-approx-N} with the values $N_{\rm i}=42$, $N_\ast=45$, and $N_{\rm f}=49$. Additionally, the initial value of ${\cal R}$ and $\Pi$ at $N_{\rm i}$ are supplied by the solutions of~\ref{eq:eom-R-Pi}. The figure distinctly illustrates that the conservation of conjugate momentum, and consequently, the equations \ref{eq:R-Pi-approx-N}, can be employed to obtain the power spectrum for the models exiting the Hubble radius before the phase where $z^2$ decreases begins. As a point of comparison, we have included the power spectrum generated using our analytical methods, assuming input values provided by the solutions~(\ref{eq:R-Pi-Hankel}). These solutions are denoted by green data points in the plot. It is important to note that this method yields results that slightly differ from the exact values. This discrepancy is expected, given that the solutions~(\ref{eq:R-Pi-Hankel}) are applicable under the assumption of a constant slow-roll parameter. To generate this plot, we selected the slow-roll parameters for each mode at the moment when the modes exit the Hubble radius.}\label{fig:PR-usr}
\end{figure}

Figure \ref{fig:PR-usr} illustrates the power spectrum of comoving curvature perturbations for the relevant scales. The blue line represents the numerically calculated power spectrum obtained by solving equation \ref{eq:H-eqns-k}. The orange data points depict the spectrum computed using \ref{eq:R-Pi-approx-N} with specific values: $N_{\rm i}=42$, $N_\ast=45$, and $N_{\rm f}=49$. The initial values of ${\cal R}$ and $\Pi$ at $\eta_{\rm i}$ are determined by the solutions of \ref{eq:eom-R-Pi}. The figure clearly demonstrates that the conservation of conjugate momentum, and consequently, the use of \ref{eq:R-Pi-approx-N}, allows us to derive the power spectrum for models exiting the Hubble radius before the onset of the phase where $z^2$ decreases. For comparison, we have included the power spectrum generated using our analytical methods, which assumes input values from the solutions given in \ref{eq:R-Pi-Hankel}. These results are represented by the green data points. For this plot, we selected the slow-roll parameters for each mode at the moment they exit the Hubble radius. It is crucial to note that this method yields results slightly deviating from the exact values. This discrepancy is anticipated, given that the solutions (\ref{eq:R-Pi-Hankel}) are applicable under the assumption of a constant slow-roll parameter.

Additionally, we recognize that, for the fastest growth in the curvature perturbation for some wavenumber is attributed to the dominant contribution from the power associated with the momentum. This also implies that the corresponding power spectrum is proportional to $k^4 \times k^3 \left| {\cal R}(\eta_{\rm i})\right|^2$. If we assume the solutions \ref{eq:R-Pi-Hankel} during the first slow-roll phase, the corresponding power spectrum has a slope of $k^{4} k^{2\epsilon/(\epsilon-1)}$. This slope is close to the steepest slope discussed in the literature (see, for example, refs. \cite{Byrnes:2018txb, Ozsoy:2019lyy} in this context).
\section{Bouncing models}\label{sec:bounce}
In this section, we demonstrate that the conservation of $\Pi$ can be utilized to compute the power spectrum after the bounce for the modes that exit the Hubble radius before the bounce. It has been known that to track scalar modes across the bounce, one must model the bouncing phase. This is typically achieved by either invoking non-canonical scalar field~\cite{Brandenberger:2012aj, Raveendran:2019idj} or employing two fields, one of which possesses negative kinetic energy~\cite{Peter:2002cn,Allen:2004vz, Raveendran:2017vfx}, or by incorporating positive spatial curvature~\cite{Martin:2003sf}. Consequently, the evolution of scalar perturbations is highly model-dependent; for instance, in the case of a two-field model, curvature perturbation can be influenced by isocurvature perturbation. In this study, we focus on the evolution of tensor perturbations, which solely depends on the evolution of the scale factor.

We shall assume that the
scale factor describing the bounce is given in terms of the
conformal time as follows~\cite{Chowdhury:2015cma,Raveendran:2017vfx}:
\begin{equation}
    a(\eta) = a_0\left[1+\left(\frac{\eta}{\eta_0}\right)^2\right]^{\frac{1}{2\left(\epsilon-1\right)}}\,,
\end{equation}
where $a_0$ is the value of the scale factor at the bounce (i.e., at $\eta= 0$), $\eta_0$ and $\epsilon$ are constant. Note that $\epsilon=3/2$ corresponds to
the specific case of matter bounce~\cite{Wands:1998yp,Brandenberger:2012zb}, while $\epsilon>3$ corresponds to an ekpyrotic bounce~\cite{Khoury:2001wf,Buchbinder:2007ad,Lehners:2007ac}.

In the context of tensor modes, the evolution equation can be written as
\begin{equation}
    h' = \frac{\Pi_{\rm h}}{a^2}\, , \quad \Pi_{\rm h}' = -a^2 \, k^2 \, h \,,\label{eq:H-eqns-k-h}
\end{equation}
where $h$ represents the amplitude of tensor perturbations, and $\Pi_{\rm h}$ is the corresponding conjugate momentum. In terms of $h$, the power spectrum is usually defined as~\cite{MUKHANOV1992203},
\begin{equation}
    \mathcal{P}_{\rm T} \equiv 8 \frac{k^3}{2 \, \pi^2} \vert h \vert ^2. \label{eq:P_T}
\end{equation}

As discussed earlier, we know that $\Pi_{\rm h}$ is conserved during the contracting phase. To calculate the value of $h$ after the bounce, we assume that $\Pi$ is conserved even slightly after the bounce, denoted by the time $\eta_{\rm f}$. We can then derive
\begin{equation}
    h(\eta_f) = h(\eta_{\rm i}) + \Pi_{\rm h}(\eta_{\rm i}) \int_{\eta_{\rm i}}^{\eta_{\rm f}} \frac{1}{a^2}  {\textrm d} \eta \,,\label{eq:h-analytical}
\end{equation}
where $\eta_{\rm i}$ is chosen such that the modes are in the super-Hubble regime, i.e., $k\, \eta \rightarrow 0$. Additionally, we have the limit $\eta_{\rm i} / \eta_0\gg 0$. In this limit, solving \ref{eq:H-eqns-k-h} on can obtain
\begin{subequations}
			\begin{align}
			h(\eta_{\rm i})&=\frac{1}{a(\eta_{\rm i}) \, \Mpl}\,\sqrt{\frac{-\eta_{\rm i}\,  \pi}{4}}\,e^{i (\nu +1/2)\pi/2} H_{\nu}^{(1)}(-k \eta_{\rm i}) ,\\
			\Pi_{\rm h}(\eta_{\rm i})&=-a(\eta_{\rm i})\,\Mpl \sqrt{\frac{-\eta_{\rm i}\,  \pi}{4}}\,ke^{i (\nu +1/2)\pi/2}  H_{\nu-1}^{(1)}(-k \eta_{\rm i})~,						    
			\end{align}\label{eq:h-Pi-Hankel}
		\end{subequations}
where $\nu \equiv \frac{1}{2} + \frac{1}{1-\epsilon}$. The subsequent step involves solving the integral in \ref{eq:h-analytical} to determine the value of $h$ at a later time $\eta_{\rm f}$. The solution to the integral in \ref{eq:h-analytical} can be expressed in terms of the hypergeometric function as
\begin{equation}
\int \frac{1}{a^2}  {\textrm d} \eta = \frac{\eta}{a_0^2}\prescript{}{2}F_1\left[\frac{1}{2}, \frac{1}{2}-\nu, \frac{3}{2},- \frac{\eta^2}{\eta_0^2}\right].
\end{equation}
Given that we are dealing with a symmetric bounce, we can choose $\eta_{\rm f} = -\eta_{\rm i}$. Additionally, we calculate the value of $h$ when the mode reaches a time after the bounce, where $\vert\eta_{\rm i}/\eta_0 \vert\gg 1$. In this limit, one can obtain that 
\begin{align}
\prescript{}{2}F_1\left[\frac{1}{2}, \frac{1}{2}-\nu, \frac{3}{2},- \frac{\eta_{\rm i}^2}{\eta_0^2}\right] \approx& \frac{1}{2 \nu}\left(\frac{\eta_{\rm i}}{\eta_0}\right)^{2\nu-1} \nonumber \\ 
&+ \frac{\sqrt{\pi}}{2}  \frac{\Gamma(-\nu)}{\Gamma(1/2-\nu)} \frac{\eta_0}{\eta_{\rm i}}\,. \label{eq:H-geometrical-limit}
\end{align}
To demonstrate the application of our method, we examine the scenario where ${1<\epsilon<3}$. It is important to note that in this case, the second term on the right-hand side of the expression \ref{eq:H-geometrical-limit} is dominant. Moreover, on the super-Hubble limit (i.e., when $k\eta_{\rm i} \ll 0$), equations \ref{eq:h-Pi-Hankel} result in
\begin{subequations}
			\begin{align}
k^{3/2}\vert h (\eta_{\rm i}) \vert & \approx \frac{k}{2\,a(\eta_{\rm i}) \, \sqrt{\pi}\, \sqrt{\epsilon}\,\Mpl}\,\Gamma(\nu) \left(\frac{k\, \eta_{\rm i}}{2}\right)^{1/2+\nu}\,,\\
\frac{\Pi (\eta_{\rm i})}{k^{3/2}} &\approx \frac{a(\eta_{\rm i})\, \sqrt{\epsilon}\,\Mpl}{k \sqrt{\pi}} \, \Gamma(1-\nu) \left(\frac{k\,\eta_{\rm i}}{2}\right)^{\nu-1/2}\,. 
			\end{align}\label{eq:h-Pi-sHubble}
		\end{subequations}
Then in the limit $\vert\eta_{\rm f}/\eta_0 \vert\gg 1$, using equations \ref{eq:H-geometrical-limit} and \ref{eq:h-Pi-sHubble} the complete solution of the \ref{eq:h-analytical} can be obtained as
\begin{equation}
 k^{3/2}\vert h (\eta_{\rm f}) \vert   \approx \frac{2^{3/2}\, \Mpl}{a_0 \, \eta_0} \left(\frac{k\, \eta_0}{2}\right)^{3/2 + \nu} \left| \frac{\Gamma(1-\nu) \Gamma(-\nu) }{\Gamma(1/2 - \nu)}\right|\,.
\end{equation}
It is interesting to note that, as implied by the above expression, the amplitude of tensor perturbation, $h$ is conserved after the bounce, as anticipated.
\begin{figure}[!t]
\centering
\includegraphics[width=0.9\linewidth]{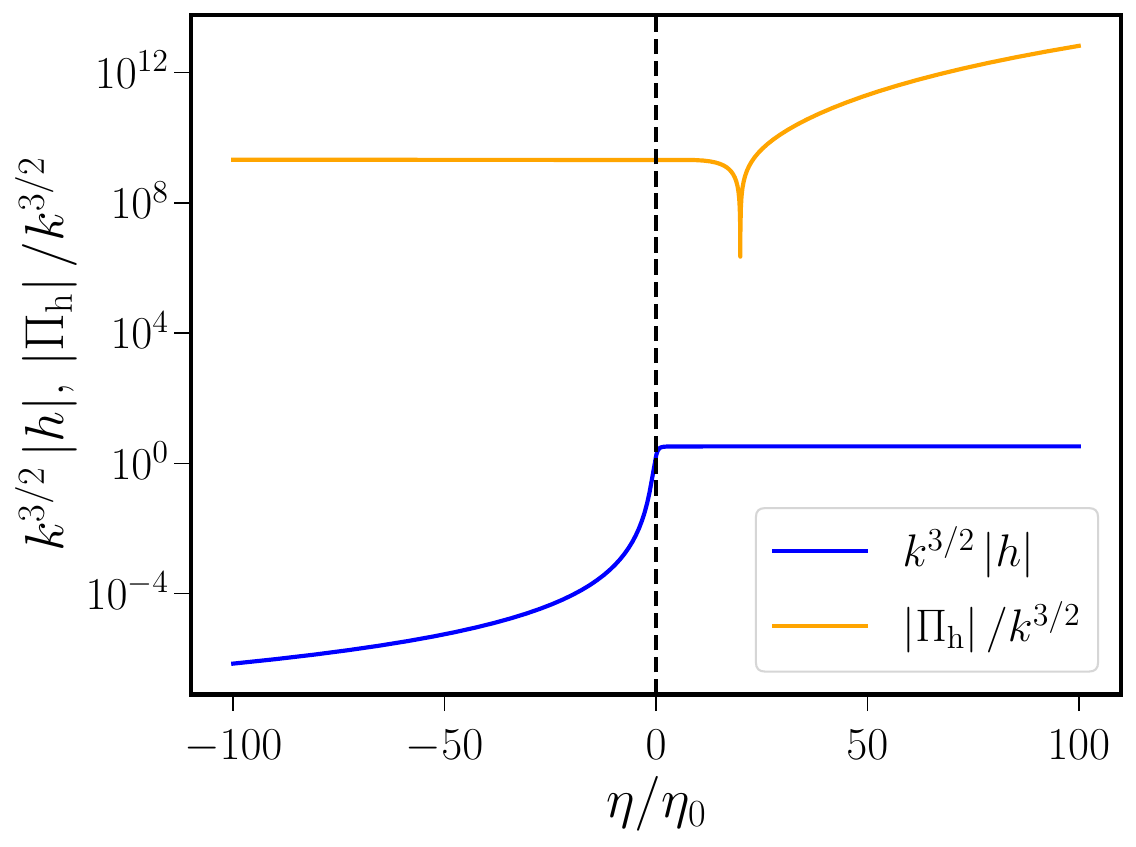}
\caption{Evolution of $\vert h \vert $ and $\vert \Pi_{\rm h}\vert $ across the bouncing phase is presented as a function of $\eta$ for $k\eta_0 = 10^{-3}$ in the matter bounce scenario, where $\epsilon=3/2$. Note that $\eta=0$ corresponds to the time of the bounce. The figure clearly shows that while $\Pi$ is conserved during the super-Hubble limit in the contracting and bouncing phases, $h$ is conserved in the expanding phase.}
\label{fig:h-P-mb}
\end{figure}
Then the power spectrum, which is defined in \ref{eq:P_T} can be obtained as
\begin{equation}
    P_{\rm T} = \frac{2^{5}\,\Mpl^2}{\left(a_0 \, \eta_0\,\pi\right)^2} \left(\frac{k\, \eta_0}{2}\right)^{3+ 2\nu} \left|\frac{\Gamma(1-\nu) \Gamma(-\nu) }{\Gamma(1/2 - \nu)}\right|^2\,.
\end{equation}

In the matter bounce scenario, where $\epsilon=3/2$, we obtain $P_{\rm T} \approx 9\,\Mpl^2/(2 a_0^2 \eta_0^2)$, which is scale-invariant as expected. This result is also consistent with the findings in~\cite{Chowdhury:2015cma}. Additionally, \ref{fig:h-P-mb} illustrates the evolution of $h$ and $\Pi$ as functions of $\eta$ in the matter bounce scenario. The figure clearly depicts that while $\Pi$ is conserved during the super-Hubble limit in the contracting and bouncing phases, $h$ is conserved in the expanding phase.
\section{Conclusions}\label{sec:conclusion}
It is well-established that the comoving curvature perturbation is conserved in slow-roll inflation on super-Hubble scales. However, it has been observed that this quantity increases during ultra-slow-roll inflation and the contracting phase of bouncing scenarios. In this work, we have demonstrated the existence of a conserved quantity, namely the conjugate momentum of comoving curvature perturbations, on super-Hubble scales in these scenarios. More specifically, we show that the conjugate momentum is conserved on super-Hubble scales during the phase when the quantity $z = a \, {\rm d}\phi/{\rm d}\log a$ decreases.

Furthermore, we illustrate that the conservation of momentum can be utilized to evaluate the evolution of the curvature perturbation during the ultra-slow-roll phase in the context of inflation, as well as during the contracting and bouncing phases in the context of bouncing scenarios. This ultimately enables obtaining an analytical estimate of the power spectrum at the end of inflation in the context of ultra-slow-roll inflation and after the bounce in the context of bouncing scenarios (see, \ref{eq:R-Pi-approx-N}). We acknowledge that the analytical expression for the power spectrum in the context of ultra-slow-roll inflation, as presented in \cite{Leach:2001zf}, is already established. Nevertheless, our study successfully replicated an approximated version of this expression, utilizing our method based on the conservation of conjugate momentum associated with the comoving curvature perturbation.

Additionally, we have examined specific models for inflation and bouncing scenarios, numerically evolved the perturbations, and showcased the conservation of conjugate momentum in these models. Additionally, we calculated the relevant power spectrum numerically and compared it with our analytical estimates.

It is interesting to identify the conserved quantities in single-field models of inflation and bouncing scenarios. The next challenge is to expand our investigation to scenarios that involve non-scalar fields or more than one scalar field, which is typically the case when modelling the bouncing phase with scalar fields \cite{Allen:2004vz, Raveendran:2017vfx}.




\acknowledgements
The author wish to thank Vincent Vennin, L. Sriramkumar, Krishnamohan Parattu, David Wands, Kazuya Koyama and Anvy Moly Tom for interesting discussions. RNR is supported by post-doctoral fellowship from the Indian Association for the Cultivation of Science, Kolkata, India.

	
	

	

	\bibliographystyle{apsrev4-1}
	\bibliography{mybibliography-cosmology}

\begin{thebibliography}{36}%
\makeatletter
\providecommand \@ifxundefined [1]{%
 \@ifx{#1\undefined}
}%
\providecommand \@ifnum [1]{%
 \ifnum #1\expandafter \@firstoftwo
 \else \expandafter \@secondoftwo
 \fi
}%
\providecommand \@ifx [1]{%
 \ifx #1\expandafter \@firstoftwo
 \else \expandafter \@secondoftwo
 \fi
}%
\providecommand \natexlab [1]{#1}%
\providecommand \enquote  [1]{``#1''}%
\providecommand \bibnamefont  [1]{#1}%
\providecommand \bibfnamefont [1]{#1}%
\providecommand \citenamefont [1]{#1}%
\providecommand \href@noop [0]{\@secondoftwo}%
\providecommand \href [0]{\begingroup \@sanitize@url \@href}%
\providecommand \@href[1]{\@@startlink{#1}\@@href}%
\providecommand \@@href[1]{\endgroup#1\@@endlink}%
\providecommand \@sanitize@url [0]{\catcode `\\12\catcode `\$12\catcode `\&12\catcode `\#12\catcode `\^12\catcode `\_12\catcode `\%12\relax}%
\providecommand \@@startlink[1]{}%
\providecommand \@@endlink[0]{}%
\providecommand \url  [0]{\begingroup\@sanitize@url \@url }%
\providecommand \@url [1]{\endgroup\@href {#1}{\urlprefix }}%
\providecommand \urlprefix  [0]{URL }%
\providecommand \Eprint [0]{\href }%
\providecommand \doibase [0]{http://dx.doi.org/}%
\providecommand \selectlanguage [0]{\@gobble}%
\providecommand \bibinfo  [0]{\@secondoftwo}%
\providecommand \bibfield  [0]{\@secondoftwo}%
\providecommand \translation [1]{[#1]}%
\providecommand \BibitemOpen [0]{}%
\providecommand \bibitemStop [0]{}%
\providecommand \bibitemNoStop [0]{.\EOS\space}%
\providecommand \EOS [0]{\spacefactor3000\relax}%
\providecommand \BibitemShut  [1]{\csname bibitem#1\endcsname}%
\let\auto@bib@innerbib\@empty
\bibitem [{\citenamefont {Ade}\ \emph {et~al.}(2016)\citenamefont {Ade} \emph {et~al.}}]{Ade:2015lrj}%
  \BibitemOpen
  \bibfield  {author} {\bibinfo {author} {\bibfnamefont {P.~A.~R.}\ \bibnamefont {Ade}} \emph {et~al.} (\bibinfo {collaboration} {Planck}),\ }\href {\doibase 10.1051/0004-6361/201525898} {\bibfield  {journal} {\bibinfo  {journal} {Astron. Astrophys.}\ }\textbf {\bibinfo {volume} {594}},\ \bibinfo {pages} {A20} (\bibinfo {year} {2016})},\ \Eprint {http://arxiv.org/abs/1502.02114} {arXiv:1502.02114 [astro-ph.CO]} \BibitemShut {NoStop}%
\bibitem [{\citenamefont {Akrami}\ \emph {et~al.}(2020)\citenamefont {Akrami} \emph {et~al.}}]{Planck:2018jri}%
  \BibitemOpen
  \bibfield  {author} {\bibinfo {author} {\bibfnamefont {Y.}~\bibnamefont {Akrami}} \emph {et~al.} (\bibinfo {collaboration} {Planck}),\ }\href {\doibase 10.1051/0004-6361/201833887} {\bibfield  {journal} {\bibinfo  {journal} {Astron. Astrophys.}\ }\textbf {\bibinfo {volume} {641}},\ \bibinfo {pages} {A10} (\bibinfo {year} {2020})},\ \Eprint {http://arxiv.org/abs/1807.06211} {arXiv:1807.06211 [astro-ph.CO]} \BibitemShut {NoStop}%
\bibitem [{\citenamefont {Mukhanov}(1988)}]{Mukhanov:1988jd}%
  \BibitemOpen
  \bibfield  {author} {\bibinfo {author} {\bibfnamefont {V.~F.}\ \bibnamefont {Mukhanov}},\ }\href@noop {} {\bibfield  {journal} {\bibinfo  {journal} {Sov. Phys. JETP}\ }\textbf {\bibinfo {volume} {67}},\ \bibinfo {pages} {1297} (\bibinfo {year} {1988})},\ \bibinfo {note} {[Zh. Eksp. Teor. Fiz.94N7,1(1988)]}\BibitemShut {NoStop}%
\bibitem [{\citenamefont {Mukhanov}\ \emph {et~al.}(1992{\natexlab{a}})\citenamefont {Mukhanov}, \citenamefont {Feldman},\ and\ \citenamefont {Brandenberger}}]{Mukhanov:1990me}%
  \BibitemOpen
  \bibfield  {author} {\bibinfo {author} {\bibfnamefont {V.~F.}\ \bibnamefont {Mukhanov}}, \bibinfo {author} {\bibfnamefont {H.~A.}\ \bibnamefont {Feldman}}, \ and\ \bibinfo {author} {\bibfnamefont {R.~H.}\ \bibnamefont {Brandenberger}},\ }\href {\doibase 10.1016/0370-1573(92)90044-Z} {\bibfield  {journal} {\bibinfo  {journal} {Phys. Rept.}\ }\textbf {\bibinfo {volume} {215}},\ \bibinfo {pages} {203} (\bibinfo {year} {1992}{\natexlab{a}})}\BibitemShut {NoStop}%
\bibitem [{\citenamefont {Martin}\ \emph {et~al.}(2014)\citenamefont {Martin}, \citenamefont {Ringeval},\ and\ \citenamefont {Vennin}}]{Martin:2013tda}%
  \BibitemOpen
  \bibfield  {author} {\bibinfo {author} {\bibfnamefont {J.}~\bibnamefont {Martin}}, \bibinfo {author} {\bibfnamefont {C.}~\bibnamefont {Ringeval}}, \ and\ \bibinfo {author} {\bibfnamefont {V.}~\bibnamefont {Vennin}},\ }\href {\doibase 10.1016/j.dark.2014.01.003} {\bibfield  {journal} {\bibinfo  {journal} {Phys. Dark Univ.}\ }\textbf {\bibinfo {volume} {5-6}},\ \bibinfo {pages} {75} (\bibinfo {year} {2014})},\ \Eprint {http://arxiv.org/abs/1303.3787} {arXiv:1303.3787 [astro-ph.CO]} \BibitemShut {NoStop}%
\bibitem [{\citenamefont {Inoue}\ and\ \citenamefont {Yokoyama}(2002)}]{Inoue:2001zt}%
  \BibitemOpen
  \bibfield  {author} {\bibinfo {author} {\bibfnamefont {S.}~\bibnamefont {Inoue}}\ and\ \bibinfo {author} {\bibfnamefont {J.}~\bibnamefont {Yokoyama}},\ }\href {\doibase 10.1016/S0370-2693(01)01369-7} {\bibfield  {journal} {\bibinfo  {journal} {Phys. Lett. B}\ }\textbf {\bibinfo {volume} {524}},\ \bibinfo {pages} {15} (\bibinfo {year} {2002})},\ \Eprint {http://arxiv.org/abs/hep-ph/0104083} {arXiv:hep-ph/0104083} \BibitemShut {NoStop}%
\bibitem [{\citenamefont {Martin}\ \emph {et~al.}(2013)\citenamefont {Martin}, \citenamefont {Motohashi},\ and\ \citenamefont {Suyama}}]{Martin:2012pe}%
  \BibitemOpen
  \bibfield  {author} {\bibinfo {author} {\bibfnamefont {J.}~\bibnamefont {Martin}}, \bibinfo {author} {\bibfnamefont {H.}~\bibnamefont {Motohashi}}, \ and\ \bibinfo {author} {\bibfnamefont {T.}~\bibnamefont {Suyama}},\ }\href {\doibase 10.1103/PhysRevD.87.023514} {\bibfield  {journal} {\bibinfo  {journal} {Phys. Rev. D}\ }\textbf {\bibinfo {volume} {87}},\ \bibinfo {pages} {023514} (\bibinfo {year} {2013})},\ \Eprint {http://arxiv.org/abs/1211.0083} {arXiv:1211.0083 [astro-ph.CO]} \BibitemShut {NoStop}%
\bibitem [{\citenamefont {Namjoo}\ \emph {et~al.}(2013)\citenamefont {Namjoo}, \citenamefont {Firouzjahi},\ and\ \citenamefont {Sasaki}}]{Namjoo:2012aa}%
  \BibitemOpen
  \bibfield  {author} {\bibinfo {author} {\bibfnamefont {M.~H.}\ \bibnamefont {Namjoo}}, \bibinfo {author} {\bibfnamefont {H.}~\bibnamefont {Firouzjahi}}, \ and\ \bibinfo {author} {\bibfnamefont {M.}~\bibnamefont {Sasaki}},\ }\href {\doibase 10.1209/0295-5075/101/39001} {\bibfield  {journal} {\bibinfo  {journal} {EPL}\ }\textbf {\bibinfo {volume} {101}},\ \bibinfo {pages} {39001} (\bibinfo {year} {2013})},\ \Eprint {http://arxiv.org/abs/1210.3692} {arXiv:1210.3692 [astro-ph.CO]} \BibitemShut {NoStop}%
\bibitem [{\citenamefont {Pattison}\ \emph {et~al.}(2018)\citenamefont {Pattison}, \citenamefont {Vennin}, \citenamefont {Assadullahi},\ and\ \citenamefont {Wands}}]{Pattison:2018bct}%
  \BibitemOpen
  \bibfield  {author} {\bibinfo {author} {\bibfnamefont {C.}~\bibnamefont {Pattison}}, \bibinfo {author} {\bibfnamefont {V.}~\bibnamefont {Vennin}}, \bibinfo {author} {\bibfnamefont {H.}~\bibnamefont {Assadullahi}}, \ and\ \bibinfo {author} {\bibfnamefont {D.}~\bibnamefont {Wands}},\ }\href {\doibase 10.1088/1475-7516/2018/08/048} {\bibfield  {journal} {\bibinfo  {journal} {JCAP}\ }\textbf {\bibinfo {volume} {08}},\ \bibinfo {pages} {048} (\bibinfo {year} {2018})},\ \Eprint {http://arxiv.org/abs/1806.09553} {arXiv:1806.09553 [astro-ph.CO]} \BibitemShut {NoStop}%
\bibitem [{\citenamefont {Garcia-Bellido}\ and\ \citenamefont {Ruiz~Morales}(2017)}]{Garcia-Bellido:2017mdw}%
  \BibitemOpen
  \bibfield  {author} {\bibinfo {author} {\bibfnamefont {J.}~\bibnamefont {Garcia-Bellido}}\ and\ \bibinfo {author} {\bibfnamefont {E.}~\bibnamefont {Ruiz~Morales}},\ }\href {\doibase 10.1016/j.dark.2017.09.007} {\bibfield  {journal} {\bibinfo  {journal} {Phys. Dark Univ.}\ }\textbf {\bibinfo {volume} {18}},\ \bibinfo {pages} {47} (\bibinfo {year} {2017})},\ \Eprint {http://arxiv.org/abs/1702.03901} {arXiv:1702.03901 [astro-ph.CO]} \BibitemShut {NoStop}%
\bibitem [{\citenamefont {Germani}\ and\ \citenamefont {Prokopec}(2017)}]{Germani:2017bcs}%
  \BibitemOpen
  \bibfield  {author} {\bibinfo {author} {\bibfnamefont {C.}~\bibnamefont {Germani}}\ and\ \bibinfo {author} {\bibfnamefont {T.}~\bibnamefont {Prokopec}},\ }\href {\doibase 10.1016/j.dark.2017.09.001} {\bibfield  {journal} {\bibinfo  {journal} {Phys. Dark Univ.}\ }\textbf {\bibinfo {volume} {18}},\ \bibinfo {pages} {6} (\bibinfo {year} {2017})},\ \Eprint {http://arxiv.org/abs/1706.04226} {arXiv:1706.04226 [astro-ph.CO]} \BibitemShut {NoStop}%
\bibitem [{\citenamefont {Ballesteros}\ and\ \citenamefont {Taoso}(2018)}]{Ballesteros:2017fsr}%
  \BibitemOpen
  \bibfield  {author} {\bibinfo {author} {\bibfnamefont {G.}~\bibnamefont {Ballesteros}}\ and\ \bibinfo {author} {\bibfnamefont {M.}~\bibnamefont {Taoso}},\ }\href {\doibase 10.1103/PhysRevD.97.023501} {\bibfield  {journal} {\bibinfo  {journal} {Phys. Rev. D}\ }\textbf {\bibinfo {volume} {97}},\ \bibinfo {pages} {023501} (\bibinfo {year} {2018})},\ \Eprint {http://arxiv.org/abs/1709.05565} {arXiv:1709.05565 [hep-ph]} \BibitemShut {NoStop}%
\bibitem [{\citenamefont {Ragavendra}\ \emph {et~al.}(2021)\citenamefont {Ragavendra}, \citenamefont {Saha}, \citenamefont {Sriramkumar},\ and\ \citenamefont {Silk}}]{Ragavendra:2020sop}%
  \BibitemOpen
  \bibfield  {author} {\bibinfo {author} {\bibfnamefont {H.~V.}\ \bibnamefont {Ragavendra}}, \bibinfo {author} {\bibfnamefont {P.}~\bibnamefont {Saha}}, \bibinfo {author} {\bibfnamefont {L.}~\bibnamefont {Sriramkumar}}, \ and\ \bibinfo {author} {\bibfnamefont {J.}~\bibnamefont {Silk}},\ }\href {\doibase 10.1103/PhysRevD.103.083510} {\bibfield  {journal} {\bibinfo  {journal} {Phys. Rev. D}\ }\textbf {\bibinfo {volume} {103}},\ \bibinfo {pages} {083510} (\bibinfo {year} {2021})},\ \Eprint {http://arxiv.org/abs/2008.12202} {arXiv:2008.12202 [astro-ph.CO]} \BibitemShut {NoStop}%
\bibitem [{\citenamefont {Ragavendra}\ and\ \citenamefont {Sriramkumar}(2023)}]{Ragavendra:2023ret}%
  \BibitemOpen
  \bibfield  {author} {\bibinfo {author} {\bibfnamefont {H.~V.}\ \bibnamefont {Ragavendra}}\ and\ \bibinfo {author} {\bibfnamefont {L.}~\bibnamefont {Sriramkumar}},\ }\href {\doibase 10.3390/galaxies11010034} {\bibfield  {journal} {\bibinfo  {journal} {Galaxies}\ }\textbf {\bibinfo {volume} {11}},\ \bibinfo {pages} {34} (\bibinfo {year} {2023})},\ \Eprint {http://arxiv.org/abs/2301.08887} {arXiv:2301.08887 [astro-ph.CO]} \BibitemShut {NoStop}%
\bibitem [{\citenamefont {Jackson}\ \emph {et~al.}(2023)\citenamefont {Jackson}, \citenamefont {Assadullahi}, \citenamefont {Gow}, \citenamefont {Koyama}, \citenamefont {Vennin},\ and\ \citenamefont {Wands}}]{Jackson:2023obv}%
  \BibitemOpen
  \bibfield  {author} {\bibinfo {author} {\bibfnamefont {J.~H.~P.}\ \bibnamefont {Jackson}}, \bibinfo {author} {\bibfnamefont {H.}~\bibnamefont {Assadullahi}}, \bibinfo {author} {\bibfnamefont {A.~D.}\ \bibnamefont {Gow}}, \bibinfo {author} {\bibfnamefont {K.}~\bibnamefont {Koyama}}, \bibinfo {author} {\bibfnamefont {V.}~\bibnamefont {Vennin}}, \ and\ \bibinfo {author} {\bibfnamefont {D.}~\bibnamefont {Wands}},\ }\href@noop {} {\  (\bibinfo {year} {2023})},\ \Eprint {http://arxiv.org/abs/2311.03281} {arXiv:2311.03281 [astro-ph.CO]} \BibitemShut {NoStop}%
\bibitem [{\citenamefont {Choudhury}\ and\ \citenamefont {Mazumdar}(2014)}]{Choudhury:2013woa}%
  \BibitemOpen
  \bibfield  {author} {\bibinfo {author} {\bibfnamefont {S.}~\bibnamefont {Choudhury}}\ and\ \bibinfo {author} {\bibfnamefont {A.}~\bibnamefont {Mazumdar}},\ }\href {\doibase 10.1016/j.physletb.2014.04.050} {\bibfield  {journal} {\bibinfo  {journal} {Phys. Lett. B}\ }\textbf {\bibinfo {volume} {733}},\ \bibinfo {pages} {270} (\bibinfo {year} {2014})},\ \Eprint {http://arxiv.org/abs/1307.5119} {arXiv:1307.5119 [astro-ph.CO]} \BibitemShut {NoStop}%
\bibitem [{\citenamefont {Khoury}\ \emph {et~al.}(2001)\citenamefont {Khoury}, \citenamefont {Ovrut}, \citenamefont {Steinhardt},\ and\ \citenamefont {Turok}}]{Khoury:2001wf}%
  \BibitemOpen
  \bibfield  {author} {\bibinfo {author} {\bibfnamefont {J.}~\bibnamefont {Khoury}}, \bibinfo {author} {\bibfnamefont {B.~A.}\ \bibnamefont {Ovrut}}, \bibinfo {author} {\bibfnamefont {P.~J.}\ \bibnamefont {Steinhardt}}, \ and\ \bibinfo {author} {\bibfnamefont {N.}~\bibnamefont {Turok}},\ }\href {\doibase 10.1103/PhysRevD.64.123522} {\bibfield  {journal} {\bibinfo  {journal} {Phys. Rev. D}\ }\textbf {\bibinfo {volume} {64}},\ \bibinfo {pages} {123522} (\bibinfo {year} {2001})},\ \Eprint {http://arxiv.org/abs/hep-th/0103239} {arXiv:hep-th/0103239} \BibitemShut {NoStop}%
\bibitem [{\citenamefont {Buchbinder}\ \emph {et~al.}(2007)\citenamefont {Buchbinder}, \citenamefont {Khoury},\ and\ \citenamefont {Ovrut}}]{Buchbinder:2007ad}%
  \BibitemOpen
  \bibfield  {author} {\bibinfo {author} {\bibfnamefont {E.~I.}\ \bibnamefont {Buchbinder}}, \bibinfo {author} {\bibfnamefont {J.}~\bibnamefont {Khoury}}, \ and\ \bibinfo {author} {\bibfnamefont {B.~A.}\ \bibnamefont {Ovrut}},\ }\href {\doibase 10.1103/PhysRevD.76.123503} {\bibfield  {journal} {\bibinfo  {journal} {Phys. Rev. D}\ }\textbf {\bibinfo {volume} {76}},\ \bibinfo {pages} {123503} (\bibinfo {year} {2007})},\ \Eprint {http://arxiv.org/abs/hep-th/0702154} {arXiv:hep-th/0702154} \BibitemShut {NoStop}%
\bibitem [{\citenamefont {Lehners}\ \emph {et~al.}(2007)\citenamefont {Lehners}, \citenamefont {McFadden}, \citenamefont {Turok},\ and\ \citenamefont {Steinhardt}}]{Lehners:2007ac}%
  \BibitemOpen
  \bibfield  {author} {\bibinfo {author} {\bibfnamefont {J.-L.}\ \bibnamefont {Lehners}}, \bibinfo {author} {\bibfnamefont {P.}~\bibnamefont {McFadden}}, \bibinfo {author} {\bibfnamefont {N.}~\bibnamefont {Turok}}, \ and\ \bibinfo {author} {\bibfnamefont {P.~J.}\ \bibnamefont {Steinhardt}},\ }\href {\doibase 10.1103/PhysRevD.76.103501} {\bibfield  {journal} {\bibinfo  {journal} {Phys. Rev. D}\ }\textbf {\bibinfo {volume} {76}},\ \bibinfo {pages} {103501} (\bibinfo {year} {2007})},\ \Eprint {http://arxiv.org/abs/hep-th/0702153} {arXiv:hep-th/0702153} \BibitemShut {NoStop}%
\bibitem [{\citenamefont {Brandenberger}\ and\ \citenamefont {Martin}(2013)}]{Brandenberger:2012aj}%
  \BibitemOpen
  \bibfield  {author} {\bibinfo {author} {\bibfnamefont {R.~H.}\ \bibnamefont {Brandenberger}}\ and\ \bibinfo {author} {\bibfnamefont {J.}~\bibnamefont {Martin}},\ }\href {\doibase 10.1088/0264-9381/30/11/113001} {\bibfield  {journal} {\bibinfo  {journal} {Class. Quant. Grav.}\ }\textbf {\bibinfo {volume} {30}},\ \bibinfo {pages} {113001} (\bibinfo {year} {2013})},\ \Eprint {http://arxiv.org/abs/1211.6753} {arXiv:1211.6753 [astro-ph.CO]} \BibitemShut {NoStop}%
\bibitem [{\citenamefont {Brandenberger}\ and\ \citenamefont {Peter}(2017)}]{Brandenberger:2016vhg}%
  \BibitemOpen
  \bibfield  {author} {\bibinfo {author} {\bibfnamefont {R.}~\bibnamefont {Brandenberger}}\ and\ \bibinfo {author} {\bibfnamefont {P.}~\bibnamefont {Peter}},\ }\href {\doibase 10.1007/s10701-016-0057-0} {\bibfield  {journal} {\bibinfo  {journal} {Found. Phys.}\ }\textbf {\bibinfo {volume} {47}},\ \bibinfo {pages} {797} (\bibinfo {year} {2017})},\ \Eprint {http://arxiv.org/abs/1603.05834} {arXiv:1603.05834 [hep-th]} \BibitemShut {NoStop}%
\bibitem [{\citenamefont {Raveendran}\ \emph {et~al.}(2018)\citenamefont {Raveendran}, \citenamefont {Chowdhury},\ and\ \citenamefont {Sriramkumar}}]{Raveendran:2017vfx}%
  \BibitemOpen
  \bibfield  {author} {\bibinfo {author} {\bibfnamefont {R.~N.}\ \bibnamefont {Raveendran}}, \bibinfo {author} {\bibfnamefont {D.}~\bibnamefont {Chowdhury}}, \ and\ \bibinfo {author} {\bibfnamefont {L.}~\bibnamefont {Sriramkumar}},\ }\href {\doibase 10.1088/1475-7516/2018/01/030} {\bibfield  {journal} {\bibinfo  {journal} {JCAP}\ }\textbf {\bibinfo {volume} {01}},\ \bibinfo {pages} {030} (\bibinfo {year} {2018})},\ \Eprint {http://arxiv.org/abs/1703.10061} {arXiv:1703.10061 [gr-qc]} \BibitemShut {NoStop}%
\bibitem [{\citenamefont {Brustein}\ \emph {et~al.}(1998)\citenamefont {Brustein}, \citenamefont {Gasperini},\ and\ \citenamefont {Veneziano}}]{Brustein:1998kq}%
  \BibitemOpen
  \bibfield  {author} {\bibinfo {author} {\bibfnamefont {R.}~\bibnamefont {Brustein}}, \bibinfo {author} {\bibfnamefont {M.}~\bibnamefont {Gasperini}}, \ and\ \bibinfo {author} {\bibfnamefont {G.}~\bibnamefont {Veneziano}},\ }\href {\doibase 10.1016/S0370-2693(98)00576-0} {\bibfield  {journal} {\bibinfo  {journal} {Phys. Lett. B}\ }\textbf {\bibinfo {volume} {431}},\ \bibinfo {pages} {277} (\bibinfo {year} {1998})},\ \Eprint {http://arxiv.org/abs/hep-th/9803018} {arXiv:hep-th/9803018} \BibitemShut {NoStop}%
\bibitem [{\citenamefont {Boyle}\ \emph {et~al.}(2004)\citenamefont {Boyle}, \citenamefont {Steinhardt},\ and\ \citenamefont {Turok}}]{Boyle:2004gv}%
  \BibitemOpen
  \bibfield  {author} {\bibinfo {author} {\bibfnamefont {L.~A.}\ \bibnamefont {Boyle}}, \bibinfo {author} {\bibfnamefont {P.~J.}\ \bibnamefont {Steinhardt}}, \ and\ \bibinfo {author} {\bibfnamefont {N.}~\bibnamefont {Turok}},\ }\href {\doibase 10.1103/PhysRevD.70.023504} {\bibfield  {journal} {\bibinfo  {journal} {Phys. Rev. D}\ }\textbf {\bibinfo {volume} {70}},\ \bibinfo {pages} {023504} (\bibinfo {year} {2004})},\ \Eprint {http://arxiv.org/abs/hep-th/0403026} {arXiv:hep-th/0403026} \BibitemShut {NoStop}%
\bibitem [{\citenamefont {Wands}(1999)}]{Wands:1998yp}%
  \BibitemOpen
  \bibfield  {author} {\bibinfo {author} {\bibfnamefont {D.}~\bibnamefont {Wands}},\ }\href {\doibase 10.1103/PhysRevD.60.023507} {\bibfield  {journal} {\bibinfo  {journal} {Phys. Rev.}\ }\textbf {\bibinfo {volume} {D60}},\ \bibinfo {pages} {023507} (\bibinfo {year} {1999})},\ \Eprint {http://arxiv.org/abs/gr-qc/9809062} {arXiv:gr-qc/9809062 [gr-qc]} \BibitemShut {NoStop}%
\bibitem [{\citenamefont {Dalianis}\ \emph {et~al.}(2019)\citenamefont {Dalianis}, \citenamefont {Kehagias},\ and\ \citenamefont {Tringas}}]{Dalianis:2018frf}%
  \BibitemOpen
  \bibfield  {author} {\bibinfo {author} {\bibfnamefont {I.}~\bibnamefont {Dalianis}}, \bibinfo {author} {\bibfnamefont {A.}~\bibnamefont {Kehagias}}, \ and\ \bibinfo {author} {\bibfnamefont {G.}~\bibnamefont {Tringas}},\ }\href {\doibase 10.1088/1475-7516/2019/01/037} {\bibfield  {journal} {\bibinfo  {journal} {JCAP}\ }\textbf {\bibinfo {volume} {01}},\ \bibinfo {pages} {037} (\bibinfo {year} {2019})},\ \Eprint {http://arxiv.org/abs/1805.09483} {arXiv:1805.09483 [astro-ph.CO]} \BibitemShut {NoStop}%
\bibitem [{\citenamefont {Leach}\ \emph {et~al.}(2001)\citenamefont {Leach}, \citenamefont {Sasaki}, \citenamefont {Wands},\ and\ \citenamefont {Liddle}}]{Leach:2001zf}%
  \BibitemOpen
  \bibfield  {author} {\bibinfo {author} {\bibfnamefont {S.~M.}\ \bibnamefont {Leach}}, \bibinfo {author} {\bibfnamefont {M.}~\bibnamefont {Sasaki}}, \bibinfo {author} {\bibfnamefont {D.}~\bibnamefont {Wands}}, \ and\ \bibinfo {author} {\bibfnamefont {A.~R.}\ \bibnamefont {Liddle}},\ }\href {\doibase 10.1103/PhysRevD.64.023512} {\bibfield  {journal} {\bibinfo  {journal} {Phys. Rev. D}\ }\textbf {\bibinfo {volume} {64}},\ \bibinfo {pages} {023512} (\bibinfo {year} {2001})},\ \Eprint {http://arxiv.org/abs/astro-ph/0101406} {arXiv:astro-ph/0101406} \BibitemShut {NoStop}%
\bibitem [{\citenamefont {Byrnes}\ \emph {et~al.}(2019)\citenamefont {Byrnes}, \citenamefont {Cole},\ and\ \citenamefont {Patil}}]{Byrnes:2018txb}%
  \BibitemOpen
  \bibfield  {author} {\bibinfo {author} {\bibfnamefont {C.~T.}\ \bibnamefont {Byrnes}}, \bibinfo {author} {\bibfnamefont {P.~S.}\ \bibnamefont {Cole}}, \ and\ \bibinfo {author} {\bibfnamefont {S.~P.}\ \bibnamefont {Patil}},\ }\href {\doibase 10.1088/1475-7516/2019/06/028} {\bibfield  {journal} {\bibinfo  {journal} {JCAP}\ }\textbf {\bibinfo {volume} {06}},\ \bibinfo {pages} {028} (\bibinfo {year} {2019})},\ \Eprint {http://arxiv.org/abs/1811.11158} {arXiv:1811.11158 [astro-ph.CO]} \BibitemShut {NoStop}%
\bibitem [{\citenamefont {\"Ozsoy}\ and\ \citenamefont {Tasinato}(2020)}]{Ozsoy:2019lyy}%
  \BibitemOpen
  \bibfield  {author} {\bibinfo {author} {\bibfnamefont {O.}~\bibnamefont {\"Ozsoy}}\ and\ \bibinfo {author} {\bibfnamefont {G.}~\bibnamefont {Tasinato}},\ }\href {\doibase 10.1088/1475-7516/2020/04/048} {\bibfield  {journal} {\bibinfo  {journal} {JCAP}\ }\textbf {\bibinfo {volume} {04}},\ \bibinfo {pages} {048} (\bibinfo {year} {2020})},\ \Eprint {http://arxiv.org/abs/1912.01061} {arXiv:1912.01061 [astro-ph.CO]} \BibitemShut {NoStop}%
\bibitem [{\citenamefont {Raveendran}(2019)}]{Raveendran:2019idj}%
  \BibitemOpen
  \bibfield  {author} {\bibinfo {author} {\bibfnamefont {R.~N.}\ \bibnamefont {Raveendran}},\ }\href {\doibase 10.1103/PhysRevD.99.103517} {\bibfield  {journal} {\bibinfo  {journal} {Phys. Rev. D}\ }\textbf {\bibinfo {volume} {99}},\ \bibinfo {pages} {103517} (\bibinfo {year} {2019})},\ \Eprint {http://arxiv.org/abs/1902.06639} {arXiv:1902.06639 [gr-qc]} \BibitemShut {NoStop}%
\bibitem [{\citenamefont {Peter}\ and\ \citenamefont {Pinto-Neto}(2002)}]{Peter:2002cn}%
  \BibitemOpen
  \bibfield  {author} {\bibinfo {author} {\bibfnamefont {P.}~\bibnamefont {Peter}}\ and\ \bibinfo {author} {\bibfnamefont {N.}~\bibnamefont {Pinto-Neto}},\ }\href {\doibase 10.1103/PhysRevD.66.063509} {\bibfield  {journal} {\bibinfo  {journal} {Phys. Rev. D}\ }\textbf {\bibinfo {volume} {66}},\ \bibinfo {pages} {063509} (\bibinfo {year} {2002})},\ \Eprint {http://arxiv.org/abs/hep-th/0203013} {arXiv:hep-th/0203013} \BibitemShut {NoStop}%
\bibitem [{\citenamefont {Allen}\ and\ \citenamefont {Wands}(2004)}]{Allen:2004vz}%
  \BibitemOpen
  \bibfield  {author} {\bibinfo {author} {\bibfnamefont {L.~E.}\ \bibnamefont {Allen}}\ and\ \bibinfo {author} {\bibfnamefont {D.}~\bibnamefont {Wands}},\ }\href {\doibase 10.1103/PhysRevD.70.063515} {\bibfield  {journal} {\bibinfo  {journal} {Phys. Rev.}\ }\textbf {\bibinfo {volume} {D70}},\ \bibinfo {pages} {063515} (\bibinfo {year} {2004})},\ \Eprint {http://arxiv.org/abs/astro-ph/0404441} {arXiv:astro-ph/0404441 [astro-ph]} \BibitemShut {NoStop}%
\bibitem [{\citenamefont {Martin}\ and\ \citenamefont {Peter}(2003)}]{Martin:2003sf}%
  \BibitemOpen
  \bibfield  {author} {\bibinfo {author} {\bibfnamefont {J.}~\bibnamefont {Martin}}\ and\ \bibinfo {author} {\bibfnamefont {P.}~\bibnamefont {Peter}},\ }\href {\doibase 10.1103/PhysRevD.68.103517} {\bibfield  {journal} {\bibinfo  {journal} {Phys. Rev. D}\ }\textbf {\bibinfo {volume} {68}},\ \bibinfo {pages} {103517} (\bibinfo {year} {2003})},\ \Eprint {http://arxiv.org/abs/hep-th/0307077} {arXiv:hep-th/0307077} \BibitemShut {NoStop}%
\bibitem [{\citenamefont {Chowdhury}\ \emph {et~al.}(2015)\citenamefont {Chowdhury}, \citenamefont {Sreenath},\ and\ \citenamefont {Sriramkumar}}]{Chowdhury:2015cma}%
  \BibitemOpen
  \bibfield  {author} {\bibinfo {author} {\bibfnamefont {D.}~\bibnamefont {Chowdhury}}, \bibinfo {author} {\bibfnamefont {V.}~\bibnamefont {Sreenath}}, \ and\ \bibinfo {author} {\bibfnamefont {L.}~\bibnamefont {Sriramkumar}},\ }\href {\doibase 10.1088/1475-7516/2015/11/002} {\bibfield  {journal} {\bibinfo  {journal} {JCAP}\ }\textbf {\bibinfo {volume} {11}},\ \bibinfo {pages} {002} (\bibinfo {year} {2015})},\ \Eprint {http://arxiv.org/abs/1506.06475} {arXiv:1506.06475 [astro-ph.CO]} \BibitemShut {NoStop}%
\bibitem [{\citenamefont {Brandenberger}(2012)}]{Brandenberger:2012zb}%
  \BibitemOpen
  \bibfield  {author} {\bibinfo {author} {\bibfnamefont {R.~H.}\ \bibnamefont {Brandenberger}},\ }\href@noop {} {\  (\bibinfo {year} {2012})},\ \Eprint {http://arxiv.org/abs/1206.4196} {arXiv:1206.4196 [astro-ph.CO]} \BibitemShut {NoStop}%
\bibitem [{\citenamefont {Mukhanov}\ \emph {et~al.}(1992{\natexlab{b}})\citenamefont {Mukhanov}, \citenamefont {Feldman},\ and\ \citenamefont {Brandenberger}}]{MUKHANOV1992203}%
  \BibitemOpen
  \bibfield  {author} {\bibinfo {author} {\bibfnamefont {V.}~\bibnamefont {Mukhanov}}, \bibinfo {author} {\bibfnamefont {H.}~\bibnamefont {Feldman}}, \ and\ \bibinfo {author} {\bibfnamefont {R.}~\bibnamefont {Brandenberger}},\ }\href {\doibase https://doi.org/10.1016/0370-1573(92)90044-Z} {\bibfield  {journal} {\bibinfo  {journal} {Physics Reports}\ }\textbf {\bibinfo {volume} {215}},\ \bibinfo {pages} {203} (\bibinfo {year} {1992}{\natexlab{b}})}\BibitemShut {NoStop}%
\end{thebibliography}%

\end{document}